\documentclass[12pt]{article}
\usepackage{amsmath}
\def\a{\alpha}
\def\b{\beta}

\def\d{\delta}

\def\g{\gamma}
\def\p{\psi}

\def\be{\begin{equation}}
\def\ee{\end{equation}}
\def\arr{\begin{array}{rll}}
\def\ea{\end{array}}
\def\bea{\begin{eqnarray}}
\def\eea{\end{eqnarray}}

\def\N2{$N{=}2$}

\def\>{\rangle}
\def\<{\langle}
\def\+{\dagger}
\def\={\ =\ }

\begin{document}
\renewcommand{\thefootnote}{\fnsymbol{footnote}}
\begin{titlepage}
\setcounter{page}{0}
\begin{flushright}
LMP-TPU--7/09  \\
\end{flushright}
\vskip 1cm
\begin{center}
{\LARGE\bf $\mathcal{N}=2$ superconformal Newton--Hooke }\\
\vskip 0.5cm
{\LARGE\bf  algebra and many--body mechanics}\\
\vskip 2cm
$
\textrm{\Large Anton Galajinsky \ }
$
\vskip 0.7cm
{\it
Laboratory of Mathematical Physics, Tomsk Polytechnic University, \\
634050 Tomsk, Lenin Ave. 30, Russian Federation} \\
{E-mail: galajin@mph.phtd.tpu.ru}

\end{center}
\vskip 1cm
\begin{abstract} \noindent
A representation of the conformal  Newton--Hooke algebra on a phase space of
$n$ particles in arbitrary dimension which
interact with one another via a generic conformal
potential and experience a universal cosmological repulsion or attraction is constructed.
The minimal $\mathcal{N}=2$ superconformal extension of the Newton--Hooke algebra
and its dynamical realization in many--body mechanics
are studied.
\end{abstract}


\vskip 0.5cm


\end{titlepage}

\renewcommand{\thefootnote}{\arabic{footnote}}
\setcounter{footnote}0

\noindent
{\bf 1. Introduction}\\
\noindent

Recent proposals for a non-relativistic version of the AdS/CFT correspondence 
\cite{son,bala}\footnote{By now there is an extensive literature on the subject. For a more complete list
of references see e.g. a recent work \cite{davo}.} stimulated renewed
interest in non--relativistic conformal (super)algebras \cite{davo}--\cite{say}
(for related earlier studies see \cite{ggt}--\cite{hu}).
Matching of symmetries in the bulk to those on the boundary is one of the principal 
ingredients of the correspondence.

There are two competing approaches to constructing non--relativistic conformal algebras.
The first option is to minimally extend the Galilei algebra by the generators of
dilatations and special conformal transformations which form the $so(1,2)$ subalgebra together with the 
generator of time translations. The resulting algebra is known as the Schr\"odinger algebra \cite{nied,hagen}. 
An alternative possibility is to consider non--relativistic
contractions of the conformal algebra
$so(d+1,2)$ (see e.g. the discussion in \cite{davo,gop,ip,mar,luk,bag}). 
This yields a larger algebra which goes under the name of conformal
Galilei algebra. Because the conformal Galilei algebra requires vanishing mass, 
the Schr\"odinger algebra is likely to have a better prospect
for quantum mechanical applications.

An analogue of the Galilei algebra in the presence of a universal cosmological repulsion or attraction is the 
Newton--Hooke algebra \cite{bac,gao,gp}. It can be derived from the (anti) de Sitter algebra by a non-relativistic 
contraction in much the same way as the Galilei algebra is obtained from the Poincar\'e algebra \cite{bac}. 
In contrast to the Galilei transformations, however, the bracket relation involving the generators of
time and space translations is modified to yield the boost: $[H,P^\a]=\pm \frac{1}{R^2} K^\a$. Here $R$ is the
radius of the parent (anti) de Sitter space.
The positive sign on the right hand side is realized in a non-relativistic spacetime with a negative cosmological 
constant.
The corresponding algebra is conventionally denoted as $nh^{-}$. The negative sign
is realized in a spacetime with a positive cosmological constant, the shorthand for the algebra being $nh^{+}$.

While in arbitrary dimension the Newton--Hooke algebra admits only one central charge,
in $(2+1)$--dimensions the second central charge is allowed which leads one to the so called exotic Newton--Hooke symmetry
\cite{gao,Olmo,Gom1,achp}.
Generalizations of the Newton--Hooke algebra associated with non--relativistic strings and branes were studied
in \cite{sak,ggk,bgk,sy}.
Various extensions of the Newton--Hooke algebra by extra vector generators and their
dynamical realizations were discussed  in \cite{luk2,Gom,tian}.

That the Newton--Hooke algebra can be extended by conformal generators was known for a long time \cite{olmo}.
However, its dynamical realization in many--body mechanics as well as supersymmetric extensions
have not yet been studied. It is natural to expect that many--body models with
superconformal Newton--Hooke symmetry may provide new insight into the non--relativistic version of
the AdS/CFT correspondence. Another motivation stems from the desire to construct new exactly solvable
many--body models in a non--relativistic spacetime with a cosmological constant and to explore novel correlations.

The purpose of this work is to construct a representation of the conformal  Newton--Hooke algebra on a phase space of
$n$ particles in arbitrary dimension which
interact with one another via a generic conformal
potential and experience a universal cosmological repulsion or attraction. We also study the minimal $\mathcal{N}=2$
superconformal extension\footnote{Recent studies of $\mathcal{N}=4$ superconformal many--body models in one dimension 
(see e.g. \cite{glp} and references therein) indicate that $\mathcal{N}=2$ is likely to be the maximal superextension 
compatible with the translation invariance
in higher dimensions.} of the Newton--Hooke algebra and provide its dynamical realization.

In Sect. 2 we discuss the minimal conformal Newton--Hooke algebra and give its dynamical realization in many--body 
mechanics.
Sect. 3 is devoted to the minimal $\mathcal{N}=2$ superconformal Newton--Hooke algebra
and its representation on a phase space of $n$ superparticles in arbitrary dimension. In each section, the negative 
and positive values of a cosmological constant
are treated separately.

\vspace{0.5cm}

\noindent
{\bf 2. Conformal extension of the Newton--Hooke algebra}\\
\noindent

The Newton--Hooke algebra describes symmetries of a non--relativistic spacetime with a cosmological 
constant \cite{bac,gao,gp}.
The generators of
time translations $H$, space translations $P^\a$, space rotations $M^{\a\b}$, and boosts $K^\a$ obey the following Lie
brackets
\bea\label{NH}
&&
[H,K^\a]=-P^\a, \quad ~ [P^\a,K^\b ]=-M \delta^{\a\b}, \quad ~ [H,P^\a]=\pm \frac{1}{R^2} K^\a,
\nonumber\\[2pt]
&&
[M^{\a\b},P^\g]=\d^{\a\g} P^\b-\d^{\b\g} P^\a, \quad ~ [M^{\a\b},K^\g]=\d^{\a\g} K^\b-\d^{\b\g} K^\a,
\nonumber\\[2pt]
&&
[M^{\a\b},M^{\g\d}]=\d^{\a\g} M^{\b\d}+\d^{\b\d} M^{\a\g}-\d^{\b\g} M^{\a\d}-\d^{\a\d} M^{\b\g},
\eea
where $M$ is the central charge and $\a=1,\dots,d$.
The algebra coincides with
the (centrally extended) Galilei algebra but for the rightmost bracket entering the first line in  (\ref{NH}).

Conformal extensions of the Newton--Hooke algebra were discussed in \cite{olmo}. Below we construct a representation of
the minimal conformal Newton--Hooke algebra on a phase space of $n$ identical particles in arbitrary dimension.

\vspace{0.5cm}

\noindent
2.1 {\it  Negative cosmological constant}\\

We first consider $nh^{-}$.
Apart form the generators displayed above, the minimal conformal extension of $nh^{-}$
involves the generators of dilatations $t_1$ and special conformal transformations $t_2$. Along with the generator
of time translations $t_0=H$ they form the $so(1,2)$ subalgebra\footnote{Redefining the generators as
$\tilde t_0=R t_0$, $\tilde t_1=2R t_1$, $\tilde t_2=2R t_2$, one gets the standard structure relations of the $so(1,2)$ 
algebra:
$[\tilde t_0,\tilde t_1]=-2 \tilde t_2$, $[\tilde t_0,\tilde t_2]=2 \tilde t_1$, $[\tilde t_1,\tilde t_2]=2 \tilde t_0$.}
\be\label{so}
[t_0,t_1]=-\frac{2}{R} t_2, \qquad [t_0,t_2]=\frac{2}{R} t_1, \qquad [t_1,t_2]=\frac{1}{2R} t_0.
\ee
Other non--vanishing Lie brackets include
\bea\label{rest}
[t_1,P^\a]=-\frac{1}{2R} P^\a, \quad [t_1,K^\a]=\frac{1}{2R} K^\a, \quad
[t_2,P^\a]=\frac{1}{2R^2} K^\a, \quad [t_2,K^\a]=\frac{1}{2} P^\a.
\eea
It is straightforward to verify that the Jacobi identities hold for the algebra determined by the structure relations
(\ref{NH}), (\ref{so}), (\ref{rest}) where $[H,P^\a]=\frac{1}{R^2} K^\a$.

In order to construct a representation of the conformal $nh^{-}$ algebra,
let us consider a
set of $n$ identical particles (of unit mass) which are parameterized by the coordinates $x^\a_i$ and 
momenta $p^\a_i$, $i=1,\dots,n$,
obeying the standard Poisson bracket
\be
\{x^\a_i,p^\b_j\}=\delta^{\a\b} \delta_{ij}.
\ee
The Hamiltonian which governs the dynamics of the system is chosen in the form
\be\label{ham}
t_0=H=\frac 12 p^\a_i p^\a_i +V(x)+\frac{1}{2 R^2} x^\a_i x^\a_i.
\ee
The last term in (\ref{ham}) is designed to describe a universal cosmological attraction \cite{gp}, while
$V(x_1,\dots,x_n)$ is supposed to be a generic conformal potential compatible with the translation and rotation invariance
\be\label{Rest}
x^\a_i \partial_{\a i} V(x)+2 V(x)=0, \qquad
\sum_{i=1}^n \partial_{\a i} V(x)=0, \qquad \left(x_i^\a \partial_{\b i}-x_i^\b
\partial_{\a i}\right)V(x)=0,
\ee
where we denoted $\partial_{\a i}=\frac{\partial}{\partial x^\a_i}$.

Guided by the previous study of the conformal algebra in the context of the oscillator potential \cite{nied1}, we
introduce the notation
\be\label{CD}
C=\frac 12 x^\a_i x^\a_i, \qquad D=-\frac 12 x^\a_i p^\a_i,
\ee
and construct the following quantities
\bea\label{ch}
&&
t_1=\frac 1R D \cos{(2t/R)} +\frac 12 H \sin{(2t/R)} -\frac{1}{R^2} C \sin{(2t/R)},
\nonumber\\[2pt]
&&
t_2=\frac{1}{R^2} C \cos{(2t/R)} -\frac 12 H \cos{(2t/R)} +\frac{1}{R} D \sin{(2t/R)} ,
\nonumber\\[2pt]
&&
P^\a= \left(\sum_{i=1}^n p^\a_i\right) \cos{(t/R)} +\frac 1R \left(\sum_{i=1}^n x^\a_i \right) \sin{(t/R)} ,
\nonumber\\[2pt]
&&
K^\a=\left(\sum_{i=1}^n x^\a_i \right)\cos{(t/R)}  -R \left(\sum_{i=1}^n p^\a_i\right) \sin{(t/R)},
\nonumber\\[2pt]
&&
M^{\a\b}=x^\a_i p^\b_i-x^\b_i p^\a_i.
\eea
Note that they explicitly depend on time. It is a matter of straightforward calculation to verify
that Eqs. (\ref{ham}), (\ref{ch}) contain the set of functions which are conserved in time and form a representation
of the conformal $nh^{-}$ algebra under the Poisson bracket. The value of the central charge for this particular 
representation is
$M=n$. For particles of mass $m$ one would have $M=nm$. So $M$ is interpreted as the mass of a system.

In order to implement the flat space limit, one first redefines the generators $t_1$ and $t_2$
so as to restore the conventional dimensions 
$t_1 \rightarrow R t_1$, $t_2 \rightarrow R^2(t_2+\frac 12 t_0)$
and then sends $R$ to infinity which corresponds to the vanishing cosmological constant. 
Taking into account the relation ${\lim}_{x\rightarrow 0} \frac{\sin{x}}{x}=1$ one 
then precisely reproduces the representation
of the Schr\"odinger algebra on a phase space of $n$ identical particles constructed in \cite{gala}.

Concluding this section we note that in \cite{ph1,nied1,bdp} (see also a recent work \cite{dhh}) 
it was demonstrated that the whole motion of a free particle 
can be associated to a half-period of the harmonic oscillator via a specific local diffeomorphism.
With the use of this transformation one can construct 
the wave function of the oscillator starting from that of the free particle.
A possibility to generalize the transformation  in \cite{ph1,nied1,bdp,dhh}
to the case of many-body conformal mechanics in the harmonic trap is an interesting open problem which we
hope to address in the near future.

\vspace{0.5cm}

\noindent
2.2 {\it   Positive cosmological constant}\\

We now turn to discuss the minimal conformal extension of  $nh^{+}$. As compared to the previous case the structure relations are slightly modified so as to take
into account the negative sign in $[H,P^\a]=-\frac{1}{R^2} K^\a$ in a way compatible with the Jacobi identities\footnote{The
redefinition $\tilde t_0=2R t_2$, $\tilde t_1=R t_0$, $\tilde t_2=2R t_1$ yields the conventional structure relations of $so(1,2)$.}
\bea
&&
[t_0,t_1]=\frac{2}{R} t_2, \qquad \quad ~ [t_0,t_2]=\frac{2}{R} t_1, \qquad \quad ~ [t_1,t_2]=-\frac{1}{2R} t_0,
\nonumber\\[2pt]
&&
[t_1,P^\a]=-\frac{1}{2R} P^\a, \quad [t_1,K^\a]=\frac{1}{2R} K^\a,
\nonumber\\[2pt]
&&
[t_2,P^\a]=\frac{1}{2R^2} K^\a, \quad ~ [t_2,K^\a]=-\frac{1}{2} P^\a.
\eea

A representation of the algebra on a phase space of $n$
identical particles is constructed by analogy with the previous case. One starts with the Hamiltonian
\be\label{ham1}
t_0=H=\frac 12 p^\a_i p^\a_i +V(x)-\frac{1}{2 R^2} x^\a_i x^\a_i,
\ee
where $V(x)$ is a generic conformal potential obeying the constraints (\ref{Rest}). The last term describes a universal
cosmological repulsion \cite{gao,gp}. It proves sufficient to replace the trigonometric functions entering (\ref{ch}) by the
hyperbolic functions and adjust the number coefficients so as to get the conserved charges
\bea
&&
t_1=\frac 1R D \cosh{(2t/R)} +\frac 12 H \sinh{(2t/R)} +\frac{1}{R^2} C \sinh{(2t/R)},
\nonumber\\[2pt]
&&
t_2=\frac{1}{R^2} C \cosh{(2t/R)} +\frac 12 H \cosh{(2t/R)} +\frac{1}{R} D \sinh{(2t/R)} ,
\nonumber\\[2pt]
&&
P^\a= \left(\sum_{i=1}^n p^\a_i\right) \cosh{(t/R)} -\frac 1R \left(\sum_{i=1}^n x^\a_i \right) \sinh{(t/R)},
\nonumber
\eea
\bea\label{ch1}
&&
K^\a=\left(\sum_{i=1}^n x^\a_i \right)\cosh{(t/R)}  -R \left(\sum_{i=1}^n p^\a_i\right) \sinh{(t/R)},
\nonumber\\[2pt]
&&
M^{\a\b}=x^\a_i p^\b_i-x^\b_i p^\a_i.
\eea
Explicit calculation then shows that they do reproduce the structure relations of
the conformal $nh^{+}$ algebra under the Poisson bracket, the corresponding central charge being $M=n$.

\vspace{0.6cm}

\noindent
{\bf 3. $\mathcal{N}=2$ superconformal extension of the Newton--Hooke algebra}\\

To the best of our knowledge, the minimal $\mathcal{N}=2$ superconformal extension of the
Newton--Hooke algebra has not yet been studied in literature.
Below we establish the structure relations of the algebra and construct a representation on a phase space of $n$
superparticles in $d$ dimensions. Depending on the sign of a cosmological constant, the analysis proceeds along
different lines. Given a spacetime with a negative cosmological constant,
our strategy is to start with reasonable supercharges. On the one hand, they should
yield a Hamiltonian which reduces to the one from the preceding section in the bosonic limit.
On the other hand, the supercharges ought to be compatible with the conformal generators.
Other generators prove to follow from the requirement of the closure of the full superalgebra.

In a spacetime with a positive cosmological constant the construction of a conventional supersymmetric extension is
problematic \cite{gao}. The problem connects to the difficulty to define conserved positive energy in the parent de Sitter space.
In this case we construct a modified superalgebra in which the bracket of two supersymmetry charges yields the conformal generator $t_2$
which is treated as a kind of a regularized Hamiltonian.

In order to accommodate $\mathcal{N}=2$ supersymmetry in many--body mechanics,
one introduces the fermionic variables $\p^\a_i$ and $\bar\p^\a_i$ which are complex conjugates of each other and obey
the brackets
\be
\{\p_i^\a,\p_j^\b \}=0, \qquad \{\bar\p_i^\a,\bar\p_j^\b \}=0,  \qquad
\{\p_i^\a,{\bar\p}_j^\b \}=-i \delta^{\a\b} \delta_{ij}.
\ee
Brackets among the bosonic and fermionic variables vanish.

\vspace{0.5cm}

\noindent
3.1 {\it  Negative cosmological constant}\\

We first consider the minimal $\mathcal{N}=2$ superconformal extension of $nh^{-}$. Our ansatz for the supersymmetry charges is
\be
Q=\p^\a_i \Big(p^\a_i+i \partial_{\a i} U(x) +\frac{i}{R} x^\a_i \Big), \qquad \bar Q=\bar\p^\a_i \Big(p^\a_i-i \partial_{\a i} U(x)-\frac{i}{R} x^\a_i \Big).
\ee
Here $U(x)$ is a prepotential which obeys the system of partial differential equations
\bea\label{Str}
\left(x_i^\a \partial_{\b i}-x_i^\b
\partial_{\a i}\right)U(x)=0, \qquad
\sum_{i=1}^n \partial_{\a i} U(x)=0,\qquad
x_i^\a \partial_{\a i} U(x)=R Z,
\eea
where $Z$ is a constant. The first two constraints in (\ref{Str}) are needed to ensure the rotation and translation invariance.
The last restriction guarantees the conformal invariance of a resulting supersymmetric mechanics. Note that these restrictions coincide with those
underling many--body quantum mechanics with $\mathcal{N}=2$ Schr\"odinger supersymmetry \cite{gm}.

Computing brackets among $Q$ and $\bar Q$
\be\label{br}
\{Q,Q\}=0, \qquad \{\bar Q,\bar Q \}=0, \qquad \{Q,\bar Q \}=-2i H,
\ee
one finds the Hamiltonian which governs the dynamics of the system
\be\label{Ham}
t_0=H=\frac{1}{2} p^\a_i p^\a_i+\frac 12 \partial_{\a i} U(x) \partial_{\a i} U(x)+
\frac{1}{2 R^2} x^\a_i x^\a_i-
\partial_{\a i} \partial_{\b j} U(x) \p^\a_i \bar\p^\b_j-\frac{1}{R} \p^\a_i \bar\p^\a_i+Z.
\ee
Because the Hamiltonian is defined up to a constant, if desirable, one can disregard
the constant contribution $Z=x^\a_i \partial_{\a i } U(x)/R$.

An immediate corollary of the structure relations
(\ref{br}) is that $Q$ and $\bar Q$ are conserved in time.
Note also that comparing the bosonic limit of (\ref{Ham}) with (\ref{ham}) one gets
\be
V(x)=\frac 12 \partial_{\a i} U(x) \partial_{\a i} U(x).
\ee
That the latter is conformal invariant is assured by the rightmost constraint in (\ref{Str}).

As the next step we modify the conformal generators in (\ref{ch}) so as to take into account the fermionic degrees of freedom
\bea\label{CG}
&&
t_1=\frac 1R D \cos{(2t/R)} +\frac 12 H \sin{(2t/R)} -\frac{1}{R^2} C \sin{(2t/R)}+\frac{1}{2R} J \sin{(2t/R)},
\nonumber\\[2pt]
&&
t_2=\frac{1}{R^2} C \cos{(2t/R)} -\frac 12 H \cos{(2t/R)} +\frac{1}{R} D \sin{(2t/R)}-\frac{1}{2R} J \cos{(2t/R)}.
\eea
Here $C$, $D$ are defined as in (\ref{CD}), $H$ is the Hamiltonian (\ref{Ham}) and
$J$ is the generator of $U(1)$ $R$--symmetry transformations which affect only the fermionic variables
\be\label{J}
J=\p^\a_i \bar\p^\a_i-RZ.
\ee
The constant contribution to $J$ might seem odd. It was included so as to avoid the appearance of a fictitious central charge
in structure relations of the superalgebra.
It is straightforward to verify that $t_1$, $t_2$ and $J$ are conserved in time. Along with $t_0=H$ they form a closed subalgebra
(vanishing brackets are omitted)
\be\label{CA}
\{ t_0,t_1\}=-\frac{2}{R} t_2, \qquad \{t_0,t_2\}=\frac{2}{R} t_1, \qquad \{t_1,t_2\}=\frac{1}{2R} t_0+\frac{1}{2 R^2} J.
\ee
Thus the $so(1,2)$ subalgebra is modified and includes the $R$--symmetry generator.

Computing brackets among the supersymmetry charges and the conformal ones (see Appendix A for explicit relations)
one gets two new generators
\be
S=\left(x^\a_i \p^\a_i +\frac{i}{2} R Q \right) e^{\frac{2it}{R}}, \qquad \bar S=\left(x^\a_i \bar\p^\a_i -\frac{i}{2} R \bar Q \right) e^{-\frac{2it}{R}},
\ee
which correspond to superconformal transformations. It is readily verified that these functions are conserved in time. However,
because they exhibit explicit dependence on time, they yield non--vanishing brackets with the Hamiltonian (see Appendix A).

As for the tensor generators, the conserved charges corresponding to translations and Galilei boosts prove to
maintain their form (see Eq. (\ref{ch}) above). The generator of rotations is to be modified so as to take into account the
fermionic degrees of freedom
\bea\label{M}
M^{\a\b}=x^\a_i p^\b_i-x^\b_i p^\a_i-i(\p^\a_i \bar\p^\b_i-\p^\b_i \bar\p^\a_i).
\eea

Computing brackets among the generators of translations and supertranslations (see Appendix A for explicit relations)
one finds two more fermionic conserved charges
\be
L^\a=\left(\sum_{i=1}^n \p^\a_i\right) e^{\frac{it}{R}}, \qquad  \bar L^\a=\left(\sum_{i=1}^n \bar\p^\a_i\right) e^{-\frac{it}{R}}.
\ee
At this stage
it is straightforward to verify that the full algebra closes and no further generators appear.
Thus one ultimately arrives at a representation of the minimal $\mathcal{N}=2$ superconformal Newton--Hooke algebra
on a phase space of $n$ superparticles in $d$ dimensions. Structure relations of the superalgebra are gathered
in Appendix A. In arbitrary dimension two central charges $M$ and $Z_1$ may enter the algebra (see Appendix A).
The representation constructed above corresponds to $M=Z_1=n$.

\vspace{0.5cm}

\noindent
3.2 {\it  Positive cosmological constant}\\

For a conformal mechanics in a space with a positive cosmological constant the energy spectrum is not bounded from below.
This is reminiscent of the fact that in the (parent) de Sitter space there does not exist conserved positive energy.
By this reason, a conventional supersymmetric extension $\{Q,\bar Q\} \sim H$ which implies $ H \ge 0$
is problematic \cite{gao}.

As was mentioned above, the conformal generator $t_2$ can be viewed as a kind of a regularized Hamiltonian. Below we demonstrate
that discarding $\{Q,\bar Q\} \sim H$ in favour of $\{Q,\bar Q\} \sim t_2$ allows one to construct
a reasonable $\mathcal{N}=2$ supersymmetric extension of the conformal $nh^{+}$ algebra.

Our starting point is the following representation of the conformal generators
\bea
&&
t_0=H=\frac{1}{2} p^\a_i p^\a_i+\frac 12 \partial_{\a i} U(x) \partial_{\a i} U(x)-
\frac{1}{2 R^2} x^\a_i x^\a_i-
\partial_{\a i} \partial_{\b j} U(x) \p^\a_i \bar\p^\b_j,
\nonumber\\[2pt]
&&
t_1=\frac 1R D \cosh{(2t/R)} +\frac 12 H \sinh{(2t/R)} +\frac{1}{R^2} C \sinh{(2t/R)},
\nonumber\\[2pt]
&&
t_2=\frac{1}{R^2} C \cosh{(2t/R)} +\frac 12 H \cosh{(2t/R)} +\frac{1}{R} D \sinh{(2t/R)}+\frac{1}{2R} J ,
\eea
where $C$, $D$ are defined as in (\ref{CD}) and
$J$ is given in (\ref{J}). It is assumed that the prepotential $U(x)$
obeys the constraints (\ref{Str}). It is readily verified that $t_1$ and $t_2$ are conserved in time.
Together with $t_0$ they form a closed subalgebra (see Appendix B).
Note that $J$ was included in $t_2$ so as to
reconcile the conformal generators with supercharges to be introduced below.

In order to fix the form of supersymmetry generators, we impose the relations
\be
\{Q,\bar Q \}=-4 i t_2, \qquad \{Q,Q \}=0, \qquad \{\bar Q, \bar Q \}=0
\ee
and demand $Q$ and $\bar Q$ to be conserved with respect to the Hamiltonian $t_0$. These requirements yield
\bea
&&
Q=\p^\a_i \Big( (p^\a_i+i \partial_{\a i} U(x)) (\cosh{(t/R)}+i \sinh{(t/R)}) -\frac{i}{R} x^\a_i(\cosh{(t/R)}-i \sinh{(t/R)})\Big),
\nonumber
\eea
\bea
&&
\bar Q=\bar\p^\a_i \Big( (p^\a_i-i \partial_{\a i} U(x)) (\cosh{(t/R)}-i \sinh{(t/R)}) +\frac{i}{R} x^\a_i(\cosh{(t/R)}+i \sinh{(t/R)})\Big).
\nonumber\\[2pt]
\eea
Superconformal generators $S$ and $\bar S$ are then determined by computing brackets among $Q$, $\bar Q$ and the conformal generators
(see Appendix B for explicit relations)
\bea
&&
S=\p^\a_i \Big( (p^\a_i+i \partial_{\a i} U(x)) (\cosh{(t/R)}-i \sinh{(t/R)}) +\frac{i}{R} x^\a_i(\cosh{(t/R)}+i \sinh{(t/R)})\Big),
\nonumber\\[2pt]
&&
\bar S=\bar\p^\a_i \Big( (p^\a_i-i \partial_{\a i} U(x)) (\cosh{(t/R)}+i \sinh{(t/R)}) -\frac{i}{R} x^\a_i(\cosh{(t/R)}-i \sinh{(t/R)})\Big).
\nonumber\\[2pt]
\eea
These are conserved in time.

It turns out that the generators of translations $P^\a$, Galilei boosts $K^\a$, and rotations $M^{\a\b}$ can be chosen as in Eqs. (\ref{ch1}) and (\ref{M}), respectively. They are conserved in time and form a closed
algebra together with other generators provided the fermionic partners of $K^\a$ are taken in the form
\bea
L^\a=\sum_{i=1}^n \p^\a_i, \qquad  \bar L^\a=\sum_{i=1}^n \bar\p^\a_i.
\eea
Because $U(x)$ is translation invariant, $L^\a$ and $\bar L^\a$ are automatically conserved in time.
The full list of structure relations of the minimal $\mathcal{N}=2$ superconformal extension of $nh^{+}$ algebra
is given in Appendix B. Like in the preceding case, two central charges $M$ and $Z_1$ enter the algebra.
The representation above corresponds to $M=Z_1=n$.

\vspace{0.6cm}

\noindent
{\bf 4. Concluding remarks}\\
\noindent

To summarize, in this work we have constructed a representation of
the minimal conformal Newton--Hooke algebra on a phase space of
$n$ particles in arbitrary dimension.
The minimal $\mathcal{N}=2$ superconformal extension
of the algebra and its dynamical realization in many--body mechanics were proposed.

Turning to possible further developments, it is interesting to construct a Lagrangian formulation
and to uncover global symmetries which correspond to the conserved charges considered above.
Then it is tempting to explore if the decoupling similarity transformation of \cite{gala,gm}
can be applied in the context of many--body quantum mechanics in a spacetime with a cosmological constant.
Finally, it is interesting to study if larger (super)conformal Newton--Hooke algebras can be derived
by non--relativistic contractions.

\vspace{0.5cm}

\noindent{\bf Acknowledgements}\\

\noindent
This work was supported in part by RF Presidential grants
MD-2590.2008.2, NS-2553.2008.2 and RFBR grant 09-02-00078.

\vspace{0.5cm}

\noindent
{\bf Appendix A}\\
\noindent

In this Appendix we display structure relations of the minimal $\mathcal{N}=2$ superconformal Newton--Hooke algebra
for the case of a negative cosmological constant. The non--vanishing graded Lie brackets read
\bea
&&
\{Q,\bar Q \}=-2i t_0, \qquad
\{Q,\bar S \}=2 R(t_2+i t_1), \quad \{\bar Q,S \}=-2R (t_2-i t_1),
\nonumber\\[2pt]
&&
\{S,\bar S \}=-\frac i2 R^2 t_0 -i R J, \quad
[t_1,Q]=\frac{i}{R^2} S, \qquad [t_2,Q]=\frac{1}{R^2} S,
\nonumber\\[2pt]
&&
[t_1,\bar Q]=-\frac{i}{R^2} \bar S,  ~ \qquad [t_2,\bar Q]=\frac{1}{R^2} \bar S, \qquad \qquad ~
[t_0,S]=\frac{2i}{R} S, \
\nonumber\\[2pt]
&&
[t_0,\bar S]=-\frac{2i}{R} \bar S, \qquad \quad
[t_1,S]=-\frac{i}{4} Q, \qquad \quad \quad [t_2,S]=\frac 14 Q,
\nonumber\\[2pt]
&&
[t_1,\bar S]=\frac i4 \bar Q, \qquad \quad \quad  [t_2,\bar S]=\frac 14 \bar Q, \qquad \qquad \quad  [Q,J]=i Q,
\nonumber\\[2pt]
&&
[\bar Q,J]=-i\bar Q, \quad \quad \quad ~~ [S,J]=i S, \qquad  \quad \qquad ~~ [\bar S,J]=-i\bar S,
\nonumber\\[2pt]
&&
[Q,P^\a]=\frac{i}{R} L^\a, \qquad \quad  [Q,K^\a]=-L^\a, \qquad \quad ~~ [\bar Q,P^\a]=-\frac{i}{R} \bar L^\a,
\nonumber\\[2pt]
&&
[\bar Q, K^\a]=-\bar L^\a, \qquad \quad  [S,P^\a]=\frac 12 L^\a, \qquad \quad ~~~~ [S, K^\a]=-\frac{i}{2} R L^\a,
\nonumber\\[2pt]
&&
[\bar S,P^\a]=\frac 12 \bar L^\a, \qquad \quad ~  [\bar S, K^\a]=\frac{i}{2} R \bar L^\a, \qquad \quad ~ [t_0,L^\a]=\frac{i}{R} L^\a,
\nonumber\\[2pt]
&&
[t_0,\bar L^\a]=-\frac{i}{R} \bar L^\a, \qquad \{Q,\bar L^\a \}=\frac{1}{R} K^\a-i P^\a, ~~ \{\bar Q,L^\a \}=-\frac{1}{R} K^\a-i P^\a,
\nonumber\\[2pt]
&&
\{ L^\a,\bar L^\b\}=-iZ_1 \delta^{\a\b}, ~ \{S,\bar L^\a \}=\frac{1}{2} R P^\a-\frac{i}{2} K^\a, ~\{\bar S, L^\a \}=-\frac 12 R P^\a-\frac{i}{2} K^\a,
\nonumber\\[2pt]
&&
[L^\a,J]=i L^\a, \qquad \qquad  [\bar L^\a,J]=-i \bar L^\a, \qquad \qquad  [t_1,P^\a]=-\frac{1}{2R} P^\a,
\nonumber
\eea
\bea
&&
[t_1, K^\a]=\frac{1}{2R} K^\a, \quad \quad ~ [t_2,P^\a]=\frac{1}{2R^2} K^\a, \quad \quad ~~  [t_2, K^\a]=\frac 12 P^\a,
\nonumber\\[2pt]
&&
[t_0,t_1]=-\frac{2}{R} t_2, \quad \quad \quad ~  [t_0,t_2]=\frac{2}{R} t_1, \qquad \qquad \quad  [t_1,t_2]=\frac{1}{2R} t_0+\frac{1}{2 R^2} J,
\nonumber\\[2pt]
&&
[t_0,K^\a]=-P^\a, \quad \qquad   [P^\a,K^\b ]=-M \delta^{\a\b}, \quad \quad ~ [t_0,P^\a]=\frac{1}{R^2} K^\a,
\nonumber\\[2pt]
&&
[M^{\a\b},P^\g]=\d^{\a\g} P^\b-\d^{\b\g} P^\a, \quad ~ [M^{\a\b},K^\g]=\d^{\a\g} K^\b-\d^{\b\g} K^\a,
\nonumber\\[2pt]
&&
[M^{\a\b},L^\g]=\d^{\a\g} L^\b-\d^{\b\g} L^\a, \quad ~ [M^{\a\b},\bar L^\g]=\d^{\a\g} \bar L^\b-\d^{\b\g} \bar L^\a,
\nonumber\\[2pt]
&&
[M^{\a\b},M^{\g\d}]=\d^{\a\g} M^{\b\d}+\d^{\b\d} M^{\a\g}-\d^{\b\g} M^{\a\d}-\d^{\a\d} M^{\b\g}.
\nonumber
\eea

\vspace{0.5cm}

\noindent
{\bf Appendix B}\\
\noindent

In this Appendix we give structure relations of the minimal $\mathcal{N}=2$ superconformal Newton--Hooke algebra
for the case of a positive cosmological constant. The non--vanishing graded Lie brackets read
\bea
&&
\{Q,\bar Q \}=-4i t_2, \qquad
\{Q,\bar S \}=-4i(\frac 12 t_0+i t_1), \quad \{\bar Q,S \}=-4i(\frac 12 t_0-i t_1),
\nonumber\\[2pt]
&&
\{S,\bar S \}=-4i (t_2-\frac 1R J), \quad
[t_1,Q]=-\frac{1}{2R} S, \qquad ~~ [t_0,Q]=\frac{i}{R} S,
\nonumber\\[2pt]
&&
[t_1,\bar Q]=-\frac{1}{2R} \bar S,  ~ \qquad [t_0,\bar Q]=-\frac{i}{R} \bar S, \qquad \qquad \quad
[t_0,S]=-\frac{i}{R} Q, \
\nonumber\\[2pt]
&&
[t_0,\bar S]=\frac{i}{R} \bar Q, \qquad \quad ~~
[t_1,S]=-\frac{1}{2R} Q, \qquad \quad \quad ~~ [t_2,S]=-\frac iR S,
\nonumber\\[2pt]
&&
[t_1,\bar S]=-\frac{1}{2R} \bar Q, \quad \quad ~ [t_2,\bar S]=\frac iR \bar S, \qquad \qquad \quad \quad [Q,J]=i Q,
\nonumber\\[2pt]
&&
[\bar Q,J]=-i\bar Q, \quad \quad \quad ~~ [S,J]=i S, \qquad  \quad \qquad \quad ~~ [\bar S,J]=-i\bar S,
\nonumber\\[2pt]
&&
[Q,P^\a]=-\frac{i}{R} L^\a, \qquad   [Q,K^\a]=-L^\a, \qquad \quad \quad ~~ [\bar Q,P^\a]=\frac{i}{R} \bar L^\a,
\nonumber\\[2pt]
&&
[\bar Q, K^\a]=-\bar L^\a, \qquad \quad  [S,P^\a]=\frac iR L^\a, \qquad \qquad ~~  [S, K^\a]=- L^\a,
\nonumber\\[2pt]
&&
[\bar S,P^\a]=-\frac iR \bar L^\a, \qquad  ~  [\bar S, K^\a]=-\bar L^\a, \qquad \qquad ~~  [t_2,L^\a]=-\frac{i}{2R} L^\a,
\nonumber\\[2pt]
&&
[t_2,\bar L^\a]=\frac{i}{2R} \bar L^\a, \qquad ~ \{Q,\bar L^\a \}=-\frac{1}{R} K^\a-i P^\a, ~ \{\bar Q,L^\a \}=\frac{1}{R} K^\a-i P^\a,
\nonumber\\[2pt]
&&
\{ L^\a,\bar L^\b\}=-iZ_1 \delta^{\a\b}, ~ \{S,\bar L^\a \}=\frac 1R K^\a-i P^\a, \quad ~ ~\{\bar S, L^\a \}=-\frac 1R K^\a-i P^\a,
\nonumber\\[2pt]
&&
[L^\a,J]=i L^\a, \qquad \qquad  [\bar L^\a,J]=-i \bar L^\a, \qquad \qquad ~~  [t_1,P^\a]=-\frac{1}{2R} P^\a,
\nonumber
\eea
\bea
&&
[t_1, K^\a]=\frac{1}{2R} K^\a, \quad \quad ~ [t_2,P^\a]=\frac{1}{2R^2} K^\a, \quad \qquad ~  [t_2, K^\a]=-\frac 12 P^\a,
\nonumber\\[2pt]
&&
[t_0,t_1]=\frac{2}{R} t_2-\frac{1}{R^2} J,  \quad   [t_0,t_2]=\frac{2}{R} t_1, \qquad \qquad \quad ~~  [t_1,t_2]=-\frac{1}{2R} t_0,
\nonumber\\[2pt]
&&
[t_0,K^\a]=-P^\a, \quad \qquad~   [P^\a,K^\b ]=-M \delta^{\a\b}, \quad \qquad  [t_0,P^\a]=-\frac{1}{R^2} K^\a,
\nonumber\\[2pt]
&&
[M^{\a\b},P^\g]=\d^{\a\g} P^\b-\d^{\b\g} P^\a, \quad ~ [M^{\a\b},K^\g]=\d^{\a\g} K^\b-\d^{\b\g} K^\a,
\nonumber\\[2pt]
&&
[M^{\a\b},L^\g]=\d^{\a\g} L^\b-\d^{\b\g} L^\a, \quad ~~ [M^{\a\b},\bar L^\g]=\d^{\a\g} \bar L^\b-\d^{\b\g} \bar L^\a,
\nonumber\\[2pt]
&&
[M^{\a\b},M^{\g\d}]=\d^{\a\g} M^{\b\d}+\d^{\b\d} M^{\a\g}-\d^{\b\g} M^{\a\d}-\d^{\a\d} M^{\b\g}.
\nonumber
\eea

\end{document}